\documentclass[preprint,aps,prd,amsmath,amssymb]{revtex4}
\usepackage{graphicx}

\begin{document}
\title{QCD condensates in ADS/QCD}
\author{Jacopo Bechi}
\email[E-mail: ]{bechi@fi.infn.it}
\affiliation{CP$^{ \bf 3}$-Origins,
%%IFK \& IMADA, University of Southern Denmark,
Campusvej 55, DK-5230 Odense M, Denmark.}
\begin{flushright}
{\it CP$^3$- Origins: 2009-16}
\end{flushright}
\begin{abstract}
This paper focuses on some issues about condensates and renormalization in AdS/QCD models. In particular we consider the consistency of the AdS/QCD approach for scale dependent quantities as the chiral condensate questioned in some recent papers and the 4D meaning of the 5D cosmological constant in a model in which the QCD is dual to a 5D gravity theory. We will be able to give some arguments that the cosmological constant is related to the QCD gluon condensate.
\end{abstract}

\maketitle

\section{Introduction}

Recently in some papers in the AdS/QCD literature (at the best of our knowledge \cite{Cherman:2008eh} was the first to ask this important question) was raised the interesting problem about the theoretical and phenomenological consistency of the treatment of scale-dependent operators in holographic models of QCD. For the sake of simplicity, following \cite{Cherman:2008eh}, we work in the hard-wall model of \cite{Erlich:2005qh} and focus on the behavior of the quark condensate $\langle\bar{q}q\rangle$ (see also \cite{Hirn:2005vk} for an interesting phenomenological treatment of the condensate in AdS/QCD). We will show that, with some very little modification of the AdS/CFT dictionary and with an appropriate 4D interpretation of the mass scales on the 5D side, there is not inconsistence. In the second part of this letter we will do some comments about the 4D meaning of the 5D cosmological constant in a model in which the QCD is dual to a 5D Einstein theory. In particular, we will argue that the cosmological constant is in relation to the gluon condensate.

%the large-N behavior of the model (see also [JUGEAU] for similar issue).

\vspace{1cm}

Now we quickly review the model that we use for the first part of the our considerations. The 5D fields are the gauge fields $A_{L,R}=A_{L,R}^{a}T^{a}$ holographically dual to the Noether currents of the QCD chiral symmetries $SU_{L}(2)\times SU_{R}(2)$ and the scalar field $\phi$ associated with the operator $\bar{q}q$. The background metric is the $AdS_{5}$ metric

\begin{equation}
ds^{2}=G_{MN}dx^{M}dx^{N}=\frac{1}{z^{2}}(\eta_{\mu\nu}dx^{\mu}dx^{\nu}-dz^{2}),\qquad \epsilon<z<z_{m},
\end{equation}

where $\eta_{\mu\nu}=\text{diag}(+1,-1,-1,-1)$, $\epsilon$ is an UV regulator ($\epsilon\rightarrow 0$ at the end of calculation) and $z_{m}$ is related to the confinament scale. The Latin indices take values $0,1,2,3,z$ and the Greek indices take values $=0,1,2,3$. Note that all is adimensional. The bulk action is

\begin{equation}
S_{5}=\int d^{5}x\sqrt{g}\text{Tr}\Big[|D\phi|^{2}+3|\phi|^{2}-\frac{1}{4g_{5}^{2}}(F_{L}^{2}+F_{R}^{2})\Big]
\end{equation}

where $D_{\mu}\phi=\partial_{\mu}\phi-iA_{L\mu}\phi+i\phi A_{R\mu}$ and $F_{MN}=\partial_{M}A_{N}-\partial_{N}A_{M}-i[A_{M},A_{N}]$.

The background profile of $\phi$ is obtained solving the classical equation of motion with the following boundary condition

\begin{equation}
\phi(\epsilon)=\epsilon M\qquad ;\qquad \phi(z_{m})=\xi.
\end{equation}

where $M=\textbf{1}m_{q}$ is the quark mass matrix.

The solution is

\begin{equation}
\phi(z)=c_{1}z^{3}+c_{2}z
\end{equation}

with

\begin{equation}
c_{1}=\frac{\xi-Mz_{m}}{z_{m}(z_{m}^{2}-\epsilon^{2})}\qquad ; \qquad c_{2}=\frac{Mz_{m}^{3}-\xi\epsilon^{2}}{z_{m}(z_{m}^{2}-\epsilon^{2})}
\end{equation}

As shown by \cite{Klebanov:1999tb} $c_{1}=\langle\bar{q}q\rangle=\sigma\textbf{1}$.

\section{The chiral condensate}

In the way we get it, the argument of \cite{Cherman:2008eh} about the possible inconsistency of the treatment of the chiral condensate in the AdS/QCD models is the following. In the holographic models it is generally assumed that $1/z$ plays the role of the renormalization scale $\mu$. The AdS/CFT dictionary relates fields on the AdS $z=0$ boundary to QCD quantities in UV. The matching to QCD is done at asymptotically high scales, where it is weakly coupled, and $\mu\rightarrow \infty$, in accordance with the identification $1/z\sim \mu$. However, in QCD, as $\mu\rightarrow\infty$ the quark mass $m_{q}$ tends to zero and the chiral condensate $\sigma$ goes to infinity. Consistency with QCD then implies that in the holographic model $m_{q}$, which is fixed on the $z=0$ boundary, should also be zero, while $\sigma$ must diverge because the left side of the GOR relation is a Renormalization Group Invariant

\begin{equation}
m_{\pi}^{2}f_{\pi}^{2}=2m_{q}\sigma
\end{equation}

This is inconsistent with the phenomenology of the model, which requires that $\sigma\neq\infty$ in order to have a finite splitting between the $\rho$ and $a_{1}$ mesons.

Now we wish to show that there are arguments because this inconsistence to disappears. Let's begin reviewing some simple aspects of the renormalization in the AdS/CFT models. For the sake of simplicity let's consider a bulk scalar field $\phi$ with mass $m_{\phi}$. The value of this field to boundary $\phi(z=\epsilon,x)$ ($\epsilon$ is the regulator dual to the 4D UV cutoff $\Lambda_{UV}$ to be taken zero at the end of the calculation)  is dual to the source of a 4D scalar operator $O$ of conformal dimension $\Delta=2+\sqrt{4+m_{\phi}^{2}}$. In others words the 4D action has the bare term $\int d^{4}x\phi(z=\epsilon,x)O(x)$. Using the AdS/CFT dictionary (and neglecting contact divergences) the interesting non-analitic term in the limit $\epsilon\rightarrow 0$ is (considering $\Delta$ non integer for example)(for a review \cite{D'Hoker:2002aw})

\begin{equation}
\langle O(x)O(y)\rangle\simeq \epsilon^{2\Delta-8}\frac{1}{|x-y|^{2\Delta}}
\end{equation}

To eliminate the cutoff dependence we can define the renormalized operator $[O]=\epsilon^{4-\Delta}O$ and

\begin{equation}
\int d^{4}x \phi O=\int d^{4}x \phi \epsilon^{\Delta-4}[O]=\int d^{4}x [\phi][O]
\end{equation}

where, therefore, the renormalized source is

\begin{equation}
[\phi](x)=\epsilon^{\Delta-4}\phi(x,z=\epsilon).
\end{equation}

The Holographic Renormalization can be done more rigorous adding counterterms on the boundary \cite{Skenderis:2002wp} and, in any case, the boundary value $\phi^{0}(x)$ of the field $\phi(x,z)$ defined by $\phi^{0}(x)=\lim_{z\rightarrow 0}z^{\Delta-4}\phi(x,z)$ plays the rule of renormalized $\epsilon$-independent source. The renormalization scale $\mu$ is arbitrary and we are taking $\mu=1$ so all is adimensional.

So $m_{q}$ and $\sigma$ are, respectively, the quark mass and the chiral condensate renormalized to the scale $\mu$ arbitrary but $\epsilon$ independent.

%Now, to adapt the AdS/CFT dictionary to the QCD some minor change are necessary.

%Because QCD is not a CFT, it can't be defined on a conformal manifold, as is the boundary of AdS at $z=0$. So we introduce a \emph{physical} cutoff $z=\epsilon_{UV}$ of order of the Planck scale.

Then we restore the right dimensionality inserting factors of $R=\mu^{-1}$ in the formulas.
%Here $R$ have \emph{not} to be necessary identified with the AdS radius but with a arbitrary scale.
%We insert also a $R^{-1}$ factor in front to the action so that the 5D fields have 4D canonical dimensions.
So we find for the boundary behavior (taking $\epsilon\ll R$)

\begin{equation}\label{anomalous}
\phi(\epsilon)=\Big(\frac{\epsilon}{R}\Big)^{3-\Delta}\phi^{0}
\end{equation}

where we have used $\epsilon$ to restored the mass dimension of $\phi(\epsilon)$ because it is dual to the source renormalized to the scale $1/\epsilon$.
%By eq.\eqref{anomalous}, if we want to interpret $\phi^{0}$ as the quark's mass renormalized to the scale $\mu=R^{-1}$ the anomalous dimension is

%\begin{equation}
%\Delta=4+\gamma_{m} \qquad\qquad\text{instead of} \qquad\qquad \Delta=3+\gamma_{m}
%\end{equation}

%how should be because the operator $\bar{q}q$ has mass dimension $d_{\bar{q}q}=3$. So the bare mass is to be taken not $\phi(\epsilon_{UV})$ but $\frac{R}{\epsilon}\phi(\epsilon_{UV})$.

%SE LA QCD è DUALE AD UNA TEORIA DI GRAVITà 5D BACKGROUND INDIPENDENT The fact that the renormalization scale is fixed by $R$ is also required for consistency by the fact that we neglect the backreaction on the metric when we calculate the propagation of a bulk field from a point on the boundary to a other point on the boundary. So we are assuming implicitly that the Euclidean 4D momentum $p_{E}<R^{-1}$ in the two points correlation functions of 4D operator.

Now we will show that, in presence of a condensate, $\mu$ have to be fixed at the same scale of the characteristic momentum of the 4D vacuum. In fact, in the chiral limit, is worth the relation

\begin{equation}
\sigma=\frac{\xi}{z_{m}(z_{m}^{2}-\epsilon^{2})}.
\end{equation}

So, taking $\epsilon\rightarrow 0$, and remembering that $\sigma$ is dual to $\langle[\bar{q}q]_{\mu}\rangle_{adim.}$ renormalized at $\mu$ (the suffix "adim." remembers us that we are considering the dimensionless operator)

\begin{equation}
\langle[\bar{q}q]_{\mu}\rangle_{adim.}=\xi\Big(\frac{R}{z_{m}}\Big)^{3}
\end{equation}

and so $\xi$ has to be interpreted as  $\langle[\bar{q}q]_{1/z_{m}}\rangle_{adim.}$, namely as the chiral condensate renormalized at the confinement scale.

In the holographic model we can calculate the pion decay constant expanding the polarization operator of the axial current $\Pi_{A}(q^{2})$, defined by
$\Pi_{A}^{\mu\nu}=(\frac{q^{\mu}q^{\nu}}{q^{2}}-g^{\mu\nu})\Pi_{A}(q^{2})=i\int d^{4}x e^{iqx}\langle 0|J_{A}^{\mu}J_{A}^{\nu}|0\rangle$, around $q^{2}=0$: $\Pi_{A}(q^{2})=\Pi_{A}(0)+q^{2}\Pi'_{A}(0)+\mathcal{O}(q^{4})$. Considering $\xi\gg 1$ (the best phenomenological value is $\xi=4$ \cite{Da Rold:2005zs}):

\begin{equation}
\Pi_{A}(0)=f_{\pi}^{2}\sim N\frac{\xi^{2/3}}{z_{m}^{2}}
\end{equation}

e thus

\begin{equation}
f_{\pi}\sim\sqrt{N}\sigma^{1/3}
\end{equation}

and, because $f_{\pi}\sim \sqrt{N}\Lambda_{\chi SB}$ (where $\Lambda_{\chi SB}$ is the scale of chiral symmetry breaking), $\sigma$ is dual at the chiral condensate $\langle\bar{q}q\rangle$ renormalized at $\Lambda_{\chi SB}$. So we can see that the scale $\mu$ have to be taken at the physical scale of the chiral symmetry breaking.

Let's consider now the holographic realization of the Renormalization Group (RG) transformations. RG transformations in AdS/CFT can be studied by using bulk diffeomorphisms that induce a dilation on the boundary metric. This is a subgroup of the conformal transformation studied in \cite{Imbimbo:1999bj}. It's known that the AdS metric, because of the second order pole, not yield a metric at the boundary $z=0$. It yields a conformal structure instead, i.e. a metric up to conformal trasformations. Namely, indicating with $G$ the AdS metric, we can define the boundary metric by

\begin{equation}
g^{0}=z^{2}G|_{z=0}
\end{equation}

but also $g^{0}=e^{w}z^{2}G|_{z=0}$, where $w$ is a function with no zeros or poles at the boundary, is good. So the AdS metric yields a conformal structure at the boundary and the dual theory is CFT \cite{Witten:1998qj} because this theory have to be indifferent to a conformal rescaling.

Here, following \cite{Skenderis:2002wp}, we will consider the trasformations,

\begin{equation}
z=z'\mu,\qquad x^{i}=x'^{i}\mu
\end{equation}

that produces the transformation $g^{0}_{\mu\nu}(x)\rightarrow \mu^{2}g^{0}_{\mu\nu}(x\mu)$.

Let's consider a scalar field $\phi(x,z)=\phi'(x',z')$, that implies (now we are working in un general $AdS_{d+1}$ space)

\begin{equation}
\phi(x,\epsilon)=\epsilon^{d-\Delta}\phi^{0}(x)=\phi'(\mu x, \mu\epsilon)=(\mu\epsilon)^{d-\Delta}\phi^{0}(\mu x)
\end{equation}

by which we have

\begin{equation}
\mu\frac{\partial}{\partial\mu}\phi^{0}(x)=-(d-\Delta)\phi^{0}(x)
\end{equation}

which is the correct RG transformation rule for  a source of an operator of dimension $\Delta$. At this point one could be puzzled because if $\phi^{0}$ transform under a dilatation it can't be renormalized at a physical scale that instead has to be taken invariant under dilatation. We can come out considering that this is fully consistent with the fact that a CFT has not physical scale and that a minor change will be necessary here to adapt the holographic construction to QCD. Because QCD is not a CFT, it can't be defined on a conformal manifold, as is the boundary of AdS at $z=0$. So we are forced to introduce a \emph{physical} cutoff $z=\epsilon_{UV}$ that can be understood of order of the Planck scale. Of course this scale can't be changed under a dilatation and it limits the diffeomorphisms that we can consider to produce a RG transformation. Only the diffeomorphisms that keep unchanged $z$ can be considered. This is consistent with the fact that in the AdS/QCD model discussed before $\phi^{0}$ had to be understood as renormalized at a physical scale about $\Lambda_{QCD}$.
%We want to note that working with a not-dynamical background there is not a necessary geometric interpretation of the scale $\mu$ as radius of the AdS space.

%ancora di più i nostri ragionamenti e perchè ci servirà dopo consideriamo la descrizione olografica delle trasformazione del gruppo di rinormalizzazione...

\section{The gluon condensate}

To obtain more grasp on the meaning of $\mu$, let's consider now an AdS/QCD model in which the QCD is dual to the 5D Einstein gravity coupled a dilaton field $\varphi$ as in \cite{Gursoy:2007cb}\cite{Gursoy:2007er}

\begin{equation}
S_{gravity}=-M_{5}^{3}\int d^{5}x\sqrt{g}\Big[ R^{(5)}+\frac{1}{2}g^{\mu\nu}\partial_{\mu}\varphi\partial_{\nu}\varphi+V(\varphi)\Big],
\end{equation}

and $V(\varphi)=\Lambda+\sum_{n=1}\varphi^{n}c_{n}$ (because the manifold has a boundary we should have to add the Gibbons-Hawking term to the action but we neglect it because it is not important for the our argument). Let's consider also a UV cutoff at $z=\epsilon$.
%This is dual to consider the effective action of the QCD with the modes with frequencies more than $\epsilon^{-1}$ integrated \emph{a la Wilson}.

We assume that $\epsilon$ is small enough so that we can neglect the contribution of the dilaton to the gravitational action respect to the contribution of the cosmological constant $\Lambda$. This approximation is justified if $\epsilon^{-1}$ corresponds to a energy at which the QCD is in a small coupling regime. This picture is dual to the QCD with a UV cutoff $\Lambda_{UV}\simeq\epsilon^{-1}$ coupled to the dynamical 4D gravity. This gravity is induced by the integration of the modes with frequencies above $\Lambda_{UV}$ and the action is

\begin{eqnarray}\label{principal}
S_{4}^{gravity}=-M_{5}^{3}\int_{0}^{\epsilon}dz\int d^{4}x\sqrt{-g}\Big[R^{(4)}(g_{\mu\nu})+\Lambda\Big]=\nonumber\\
=-M_{5}^{3}\epsilon\int d^{4}x \sqrt{-g}\Big[R^{(4)}(g_{\mu\nu})+\Lambda\Big]
\end{eqnarray}

where $M_{5}$ is the 5D Planck scale and $S_{4}^{gravity}$ is obtained by substituting $g_{zz}(x,z)=g_{\mu z}(x,z)=0$ and $g_{\mu\nu}(x,z)=g_{\mu\nu}(x,\epsilon)=g_{\mu\nu}(x)$ into the 5D Einstein action. The dimensional reduction is justified because the holographic dictionary tell to us that the 5D modes localized at $z=\epsilon$ have energy $E \simeq \epsilon^{-1}$ (in order that to make this argument fully consistent we have to take $\epsilon\leq R$ where now $R$ is the AdS radius. Otherwise the warping effects could not be neglected). The 4D effective induced Planck scale is $M_{4}^{2}=M_{5}^{3}\epsilon$. In this effective theory the 4D gravity has a cosmological constant term which is naturally understood as the cosmological constant induced by the QCD scale anomaly (see \cite{Adler:1982ri} for a review on an early approach at some of this ideas in an pure Quantum Field Theory context).

%\begin{equation}
%\Lambda\simeq \frac{1}{M_{4}^{2}}\langle \theta_{\mu}^{\mu}\rangle
%\end{equation}

%where $\theta_{\mu}^{\mu}$ is the trace of the stress tensor of the QCD. ADLERADLERADLER.
%This interpretation of the cosmological constant seems to us right because the boundary of the AdS space, dual to the 4D space-time on which is formulated the 4D field theory, is flat. So $\Lambda$ is not to be interpreted as an external cosmological constant. This maybe also suggests that a theory without scale anomaly can't be dual at a 5D theory with dynamical background. For example SYM $\mathcal{N}=4$, that is conformal, is dual to a string theory on a fixed background. E IL CONTRARIO?????NON PUO' ESSERE VERO SENNO' VANNO A PUTTANE TUTTI I MODELLI NON CONFORMI!!

Generally also the perturbative vacuum energy contributes to the cosmological constant which a contribution proportional to the cutoff scale. As we have already seen, a change in the renormalization scale is dual to a 5D diffeomorphism that induces a Weyl transformation of the boundary metric. Because $\Lambda$ is diffeomorphism invariant, we associate this with a renormalization group invariant quantity.

To value the induced cosmological constant, let's consider the 4D semiclassical Einstein equation, with cosmological constant, that follows by a variational principle around a solution of the equation of motion

\begin{equation}\label{einstein}
\frac{\delta S_{grav.}}{\delta g^{\mu\nu}}\delta g^{\mu\nu}+\frac{\delta W_{R}}{\delta g^{\mu\nu}}\delta g^{\mu\nu}=0
\end{equation}

where $S_{grav.}$ is the 4D Einstein-Hilbert action, $W_{R}$ is the renormalized connected correlation functions functional generator of the QCD and $\delta g^{\mu\nu}$ is an arbitrary variation of the metric. Following \cite{Landau} we get

\begin{equation}
\frac{\delta S_{grav.}}{\delta g^{\mu\nu}}\delta g^{\mu\nu} = -M_{4}^{2}\int \Big(R_{\mu\nu}-\frac{1}{2}g_{\mu\nu}R-\frac{1}{2}\Lambda g_{\mu\nu}\Big)\delta g^{\mu\nu}\sqrt{-g}d^{4}x
\end{equation}

and the overall sign minus is fixed by the demand that the kinetic energy be positive.

Let's consider the following transformation of the metric (with $\sigma \ll 1$),

\begin{equation}
g^{\mu\nu}\rightarrow e^{-2\sigma}g^{\mu\nu}\simeq g^{\mu\nu}-2\sigma g^{\mu\nu}=g^{\mu\nu}+\delta g^{\mu\nu}.
\end{equation}

This transformation is equivalent to a dilatation, so is equivalent to a shift of the renormalization scale $\mu$,

\begin{equation}
\mu\rightarrow \mu+\sigma\mu=\mu+\delta \mu
\end{equation}

that implies a change of the renormalized coupling constant

\begin{equation}
g(\mu)\rightarrow g(\mu+\sigma\mu)\simeq g(\mu)+\sigma \beta \qquad\text{and}\qquad \beta\equiv\frac{\partial g}{\partial\mu}\mu
\end{equation}

Because the QCD (we neglect the quarks) is classically scale invariant, this change in the renormalized coupling constant is the only effect \footnote{Because we are working on a general metric there should be also a contribution from the gravitational Weyl Anomaly \cite{Duff:1993wm}. Because we are interested to specialize to a flat metric at the end we neglect this. However, this contribution should be $1/N$ suppressed respect to \eqref{gigi}}  that contributes to the variation of $W_{R}$

\begin{equation}\label{gigi}
\delta W_{R}=\frac{\partial W_{R}}{\partial g}\frac{\partial g}{\partial \mu}\delta \mu
\end{equation}

Starting from

\begin{equation}
e^{iW_{R}(g)}=\int DA_{\mu}e^{-\frac{i}{2g^{2}}\int\text{Tr}F^{2}\sqrt{-g}d^{4}x}
\end{equation}

we get

\begin{eqnarray}
&&\int DA_{\mu}e^{-\frac{i}{2(g+\delta g)^{2}}\int\text{Tr}F^{2}\sqrt{-g}d^{4}x}\simeq
\int DA_{\mu}e^{-\frac{i}{2g^{2}}(1-2\delta g/g)\int\text{Tr}F^{2}\sqrt{-g}d^{4}x}\simeq\nonumber\\
&&\simeq \int DA_{\mu}\Big(1+i\frac{\delta g}{g^{3}}\int\text{Tr}F^{2}\sqrt{-g}d^{4}x \Big)e^{-\frac{i}{2g^{2}}\int\text{Tr}F^{2}\sqrt{-g}d^{4}x}=\nonumber\\&&=
e^{iW_{R}(g)}\Big(1+i\frac{\partial W_{R}}{\partial g}\delta g\Big)
\end{eqnarray}

from which follows

\begin{equation}
\frac{\delta W_{R}}{\delta g}=\frac{1}{g^{3}}\int\langle \text{Tr}F^{2}\rangle \sqrt{-g}d^{4}x
\end{equation}

Let's consider, then, $\delta S_{grav.}$ under the same variation of the metric

\begin{eqnarray}
\delta S_{grav.}&=&-M_{4}^{2}\int \Big(R_{\mu\nu}-\frac{1}{2}g_{\mu\nu}R-\frac{1}{2}\Lambda g_{\mu\nu}\Big)\delta g^{\mu\nu}\sqrt{-g}d^{4}x=\nonumber\\
&=& M_{4}^{2}\int \Big(R_{\mu\nu}-\frac{1}{2}g_{\mu\nu}R-\frac{1}{2}\Lambda g_{\mu\nu}\Big) 2\sigma g^{\mu\nu}\sqrt{-g}d^{4}x=\nonumber\\
&=&-M_{4}^{2}\int 2\sigma\Big(R+2\Lambda\Big) \sqrt{-g}d^{4}x
\end{eqnarray}

Putting everything together

\begin{equation}\label{einsemi}
\delta S_{grav.}+\delta W=-\int 2\sigma M_{4}^{2} \Big(R+2\Lambda\Big)\sqrt{-g}d^{4}x+\int \sigma \frac{\beta}{g^{3}}\langle \text{Tr}F^{2}\rangle\sqrt{-g} d^{4}x=0
\end{equation}

Because we are interested to the induced cosmological constant $\Lambda_{ind.}$, let's specialize \eqref{einsemi} to the case of flat metric and, using the translational invariance,

%and so

%\begin{equation}\label{einsemi}
%R+\Lambda=\frac{1}{M_{4}^{2}}\frac{\beta}{4g^{3}}\langle\text{Tr}F^{2}\rangle
%\end{equation}

%Because we are interested to the induced cosmological constant $\Lambda_{ind.}$, specialize \eqref{einsemi} to the case of flat metric

\begin{equation}
\Lambda=\frac{1}{M_{4}^{2}}\frac{\beta}{4g^{3}}\langle\text{Tr}F^{2}\rangle
\end{equation}

and so the induced cosmological constant is

\begin{equation}
\Lambda_{ind.}=-\frac{1}{M_{4}^{2}}\frac{\beta}{4g^{3}}\langle\text{Tr}F^{2}\rangle
\end{equation}

%At this point is worth pause to say that to get \eqref{einsemi} we have taken the approximation that the QFT is on a flat space-time. In other terms, the QFT effects induce the metric curvature but we neglect the backreaction of the gravitational field on the QFT. This amount to neglect a term of higher order in $M_{4}^{-1}$, and, as we will see later, this means to neglect terms of higher order in $1/N$.

%Using \eqref{einsemi} we can match the cosmological constant of the 5D model with a 4D quantity:

%\begin{equation}
%\Lambda=-\frac{1}{M_{4}^{2}}\frac{\beta}{4g^{3}}\langle \text{Tr} F^{2}\rangle
%\end{equation}

%Because we are looking a region of energies in which the QCD is in a perturbative regime, we take the approximation

%\begin{equation}
%\beta(g)\simeq -b_{0}g^{3}/(4\pi)^{2}\qquad b_{0}=\frac{11}{3}N
%\end{equation}

where the condensate is valued on a flat background and, because $\beta<0$ by asymptotic freedom, the sign of $\Lambda$ depends by the sign of the gluon condensate. This quantity gets contribution only by non perturbative field configurations that in the semiclassical region at which we are looking are dominated by the instantons. The gluon condensate can be write as $\langle \text{Tr} F^{2}\rangle=2(\textbf{B}^{2}-\textbf{E}^{2})$, where $\textbf{B}$ and $\textbf{E}$ are, respectively, the cromomagnetic and cromoelectric fields (see e.g. \cite{Diakonov:2002fq}). Indeed, instanton is a tunneling processes, it occurs in imaginary time; therefore in Minkowski space one has imaginary cromoelectric fields and so the gluon condensate is positive.

From this consideration, it follows that $\Lambda>0$ in \eqref{principal}, and this is the sign that we have to take to get AdS space. It's nice that the necessity of a asymptotic AdS behavior in 5D is linked to the presence of the instantons in the QCD vacuum. Moreover, from it, we can get hints about the origin of the constant part of the dilaton potential in more stringy model as \cite{Gursoy:2007cb}\cite{Gursoy:2007er}; if our analysis is right, the constant part of the potential have to be generated by non perturbative effects dual to tunneling events in QCD.

In our derivation, we have used an arbitrary cutoff $\epsilon$ and, keeping it explicit, we have

\begin{equation}
\Lambda=\frac{1}{M_{5}^{3}\epsilon}\frac{\beta}{4g^{3}}\langle\text{Tr}F^{2}\rangle,
\end{equation}

Let's consider the case $\epsilon=R$, where $R$ is the AdS radius; using $M_{5}^{3}R^{3}\sim N^{2}$ (see e.g. \cite{ArkaniHamed:2000ds}), $\Lambda=12/R^{2}$ and the fact that $\frac{\beta}{4g^{3}}\langle\text{Tr}F^{2}\rangle$ is a Renormalization Group Invariant and so it will be (see the end of this section),

\begin{equation}
\frac{\beta}{4g^{3}}\langle\text{Tr}F^{2}\rangle\simeq c \Lambda_{QCD}^{4}\qquad \text{with}\qquad c\sim O(N^{2}).
\end{equation}

we find

\begin{equation}
\frac{1}{R}\sim \Lambda_{QCD}
\end{equation}

%but $\frac{\beta}{4g^{3}}\langle\text{Tr}F^{2}\rangle$ is a Renormalization Group invariant (RGI) and so it will be

%\begin{equation}
%\frac{\beta}{4g^{3}}\langle\text{Tr}F^{2}\rangle\simeq c \Lambda_{QCD}^{4}\qquad \text{with}\qquad c\sim O(N^{2}).
%\end{equation}

%Using $M_{5}^{3}R^{3}\sim N^{2}$ (see e.g. [PORRATI-RANDALL]) and $\Lambda=12/R^{2}$, we find

Moreover, if we consider a generic $\epsilon$, we get

\begin{equation}\label{evans}
\frac{1}{\epsilon}\sim 10 \frac{1}{R}
\end{equation}

The numerical factor in \eqref{evans} is only indicative and it should not to be taken too seriously; \eqref{evans} only says that the UV cutoff is to be taken about to $\Lambda_{QCD}$. So this seems to indicate that AdS/QCD have to be understood as a low-energy description valid only below some cutoff scale which is generally a few GeV. It is in accord to the fact that there are many things AdS/QCD gets wrong at high energy \cite{Hofman:2008ar}\cite{Csaki:2008dt} (see also \cite{Evans:2006ea} where an UV cutoff at $\sim$ GeV is imposed to improve the phenomenological analysis).

Now one could wonder that in the $N$ counting we have taken $\langle \text{Tr} (F_{\mu\nu})^{2}\rangle\sim N$ like follows by Feynman diagram arguments and not $\langle \text{Tr} (F_{\mu\nu})^{2}\rangle\sim e^{-N}$ as is for a gas of semiclassical objects. Several facts, the most important is maybe the success of the Instantons Liquid Model (for a review \cite{Schafer:1996wv}), suggest that in the QCD vacuum the interactions between instantons are important and the instanton ensemble is more similar to a liquid than to a gas. Schafer \cite{Schafer:2004gy} has shown that, in the Instanton Liquid picture, the instanton density in the large $N$ limit is proportional to $N$. This is also in accord with lattice results.

By the point of view of this section, the fact that in presence of quarks, the fermion condensate scale is $\Lambda_{\chi SB}\sim 1/R$ is an obvious fact. Moreover, the identification of $\Lambda$ with the non perturbative gluon effects lets to us to understand why we can neglect the backreaction of the fermion condensate on the AdS metric and still to get good results. This is the same approximation that work successfully in SVZ \cite{Shifman:1978bx}\cite{Shifman:1978by}.

\section{Conclusions}

In this paper we have studied some issues about the condensates in two different models of holographic QCD.
We have shown that the treatment of the chiral condensate is consistent in AdS/QCD models and that the renormalization scale have to be taken at the scale of the condensate. Moreover, we have found a relation between the cosmological constant and the gluon condensate in a holographic model with dynamical background. This relation indicates that the cosmological constant of the 5D gravity theory dual to the QCD is related at the instanton physics in 4D. Finally, also in this model appears that the UV cutoff of the AdS/QCD models have to be taken at a few GeV in accord a some precedent analysis \cite{Evans:2006ea}.

We think that a better understanding of these questions should be useful to explain why certain models work especially well and to obtain more grasp on the possibilities and the limitations of the AdS/QCD approach.

\textbf{Note:} when this work was completed we noted \cite{Haba:2008nz} that does some similar consideration about the renormalization of the chiral condensate in an Conformal Technicolor/Holographic context.

\vspace{1cm}

\textbf{Acknowlegdements}: I would like thank the $CP^{3}-Origins$ for hospitality and the Fondazione Angelo della Riccia for financial support.

\end{document}